\documentclass[twocolumn,aps,pra,tightenlines,amsmath,amssymb]{revtex4-1}
\usepackage{graphicx}
\usepackage{amssymb}
\usepackage{dcolumn}
\usepackage{amsmath}
\usepackage{bm}
\usepackage{colordvi}

\begin{document}

\title{Correlation measures in bipartite states and entanglement irreversibility}
\author{Shengjun Wu}

\affiliation{Hefei National Laboratory for Physical Sciences at Microscale and Department of Modern Physics, University of Science and Technology of China, Hefei, Anhui 230026, China \\
Institute for Quantum Information Science, University of Calgary, Calgary, Alberta T2N 1N4, Canada }

\date{\today}

\begin{abstract}

We derive quantitative relations among several naturally defined measures of classical and nonclassical correlations in a bipartite quantum state.
We also obtain an upper bound of entanglement irreversibility and a sufficient condition for reversible entanglement.  The additivity of entanglement of formation is directly related to the additivity of quantum discord as well as a certain measure of classical correlation.

\end{abstract}


\maketitle

\section{Introduction}

The study of nonclassical (quantum) correlation can be traced back to the 1930s, when Schr{\"o}dinger \cite{Sc} invented the word {\sl entanglement} to describe the inseparability of our knowledge of a composite system, and Einstein \cite{EPR} used this peculiar correlation to argue that quantum mechanics is not complete.
A formal mathematical definition of entanglement arises from the point of view that entanglement cannot be created locally even with classical communication, therefore entangled states are those that cannot be written as a convex sum of product states.  This essentially motivated the definition of entanglement of formation $E^F$ and entanglement cost $E^C$, the latter as the asymptotic version of the former.  Various other measures, such as entanglement of distillation $E^D$ \cite{BBPS96,Bennett96hash}, relative entropy of entanglement \cite{VPRK97}, squashed entanglement \cite{CW04}, arise to describe different properties of entangled states.

However, entanglement (in the strict sense, nonzero entanglement of formation) is not the only aspect of quantum (nonclassical) correlation. Quantum discord $\mathcal{Q}^d$, formulated by Ollivier and Zurek \cite{OZ01}, describes a different aspect of nonclassical correlation.
Other measures of nonclassical correlation beyond entanglement include the quantum deficit \cite{Oppenheim2002,Horodecki2003a}, measurement-induced disturbance \cite{Luo2008},
symmetric discord \cite{Piani2008,wu2009correl}, relative entropy of discord and dissonance \cite{Modi2010}, geometric discord \cite{Dakic2010,LuoFu2010}, and continuous-variable discord \cite{AD2010,GP2010}. For a nice review of the different measures of quantum correlation beyond entanglement, see \cite{Modi2011}.
Separable states could in general, have nonzero quantum correlation; and the only states that have zero quantum discord are those classical-quantum (CQ) states and the only states that have zero symmetric discord are those classical-classical (CC) states. Both quantum discord and symmetric discord have been studied in DQC1 model of quantum computation \cite{KL98}, and it is widely believed that the speedup in quantum computation could be attributed to the existence of nonclassical correlation beyond entanglement \cite{DSC08,wu2009correl}.

Despite the enormous works on quantifying classical and quantum correlations, there are many essential issues that need to be solved or understood, such as the differences in various measures of classical and quantum correlations, the irreversibility in entanglement manipulation, and the additivity problem of entanglement of formation. The purpose of this article is to establish quantitative relations among several entropic measures of classical and nonclassical correlations, especially those related to quantum discord and the symmetric discord, and provide insights into the irreversibility of entanglement manipulation and additivity of various measures. In this article, the measures of classical and nonclassical correlations are defined according to optimizations over positive-operator valued measure (POVM) measurements, while most discussions in the literature are based on projective measurements.

The article is arranged as follows.  In section II, we start with an introduction to several naturally defined measures of classical and nonclassical correlations. In section III, we establish inequality relations for several different measures of classical and nonclassical correlations. We use these relations to investigate some open questions in the following two sections. In section IV,  we obtain an upper bound of entanglement irreversibility and a sufficient condition for reversible entanglement. In section V, we address the additivity problem and connect additivity of entanglement of formation to that of quantum discord as well as a certain measure of classical correlation.

\section{Measures of classical and nonclassical correlations}

For a bipartite quantum state $\rho_{AB}$, quantum discord $\mathcal{Q}^d_{\bar{A}B}$ is defined as \cite{OZ01}
\begin{equation}
\mathcal{Q}^d_{\bar{A}B} (\rho_{AB})  = S(A:B) - \mathcal{C}^h_{\bar{A}B} .
\end{equation}
Here $S(A:B)=S(\rho_A) +S(\rho_B) - S(\rho_{AB})$ is the quantum mutual information with $S(\cdot)$ denoting the von Neumann entropy.  $\mathcal{C}^h_{\bar{A}B}$ is a measure of classical correlation defined as \cite{HV2001}
\begin{equation}
\mathcal{C}^h_{\bar{A}B} (\rho_{AB}) = \max_{ \{ \Pi _i \} } [ S(\rho_B) - \sum_i p_i S(\rho_i^B) ]
\end{equation}
where $p_i = tr_{AB} ((\Pi_i \otimes I_B) \rho_{AB})$ denotes the probability of obtaining the $i$-th result for a POVM measurement $\{\Pi_i \}$ on system A, and $\rho_i^B = tr_{A} ((\Pi_i \otimes I_B) \rho_{AB})/p_i$ denotes the state of system B conditional on Alice's $i$-th measurement result. $\mathcal{C}^h_{\bar{A}B}$ can be viewed as the Holevo bound of the ensemble $\{p_i; \rho_i^B \}$ that is prepared for Bob by Alice via her local measurement, hence the superscript $h$ is adopted.  The bar on $A$ in both $\mathcal{Q}^d_{\bar{A}B}$ and $\mathcal{C}^h_{\bar{A}B} $ reminds us that the measurement is performed on system A, as they are asymmetrical in general.

There are alternative measures of classical correlation in a bipartite state $\rho_{AB}$. Suppose Alice and Bob can perform any local POVM measurements $\{\Pi_i^A \otimes \Pi_j^B \}$ (with $\Pi_i^A \geq 0$, $\Pi_i^B \geq 0$, $\sum_i \Pi_i^A = I_A$ and $\sum_j \Pi_j^B = I_B$), from the joint probability distribution $p_{ij}=tr_{AB}(\Pi_i^A \otimes \Pi_j^B  \rho_{AB})$ one can define a classical mutual information
\begin{equation}
I(A:B)(p_{ij})  = H\{ p_i^A \} +H\{ p_j^B \} -H\{ p_{ij} \}
\end{equation}
where $H\{ \cdot \}$ denotes the Shannon entropy of the corresponding probability distribution, $p_i^A$ and $p_j^B$ are marginal probability distributions of $p_{ij}$.
Let $\mathcal{C}_{\bar{A}\bar{B}}$ denote the maximum classical correlation that can be extracted by local measurements, i.e.,
\begin{equation}
\mathcal{C}_{\bar{A}\bar{B}} (\rho_{AB}) = \max_{ \{ \Pi_i^A \otimes \Pi_j^B \} } I(A:B) (p_{ij}).
\end{equation}
$\mathcal{C}_{\bar{A}\bar{B}}$ is a natural measure of the maximum amount of classical correlation accessible by means of local measurements, and it is symmetrical with respect to both systems. A symmetric measure of nonclassical correlation was proposed and discussed in detail by Wu, Poulsen, and M{\o}lmer in \cite{wu2009correl}, and it is referred to as the WPM measure or the symmetric discord in the literature \cite{LCS11}. The symmetric discord is given by the difference of quantum mutual information and $\mathcal{C}_{\bar{A}\bar{B}}$:
\begin{equation}
\mathcal{Q}^s_{AB}(\rho_{AB})  = S(A:B) - \mathcal{C}_{\bar{A}\bar{B}}(\rho_{AB}) .
\end{equation}
Since $\mathcal{C}^h_{\bar{A}B}$ can be viewed as the Holevo bound of ensemble $\{p_i; \rho_i^B \}$ prepared for Bob by Alice's measurement on her system A, it is an upper bound of the accessible information $\mathcal{C}_{\bar{A}\bar{B}}$ via a subsequent measurement by Bob on his system B \cite{wu2009correl}, i.e.,
\begin{equation}
\mathcal{C}^h_{\bar{A}B} \geq \mathcal{C}_{\bar{A}\bar{B}} .
\end{equation}
Similarly,
\begin{equation}
\mathcal{C}^h_{A\bar{B}} \geq \mathcal{C}_{\bar{A}\bar{B}} .
\end{equation}
Therefore, we have
\begin{eqnarray}
\mathcal{Q}^d_{\bar{A}B} &\leq & \mathcal{Q}^s_{AB} \\
\mathcal{Q}^d_{A\bar{B}} &\leq & \mathcal{Q}^s_{AB} .
\end{eqnarray}
For a review of entropic measures of nonclassical correlation, see \cite{LCS11}.

If classical communication is also allowed in addition to the local measurements, other alternative measures of classical correlation are possible. Suppose Alice and Bob can perform any local operations and one-way classical communication (LOCC), then an alternative measure $\mathcal{C}^l_{\vec{AB}}$ of classical correlation in $\rho_{AB}$ is defined as the maximum classical correlation gain
\begin{equation}
\mathcal{C}^l_{\vec{AB}} (\rho_{AB}) = \max_{ \mathcal{E}} [I(A:B)(\mathcal{E})-ccc(\mathcal{E})]
\end{equation}
where $I(A:B)(\mathcal{E})$ denotes the maximum classical mutual information that could be established via the LOCC protocol $\mathcal{E}$, and $ccc(\mathcal{E})$ denotes the amount of classical communication cost, the arrow in $\vec{AB}$ reminds us that only one-way classical communication from Alice to Bob is allowed in the LOCC protocol $\mathcal{E}$. Similar quantities can be defined if other protocols such as two-way communication are allowed.  Since $\mathcal{C}_{\bar{A}\bar{B}}$ can be viewed as $\mathcal{C}^l_{\vec{AB}}$ with zero communication, we have
\begin{equation}
\mathcal{C}_{\bar{A}\bar{B}} \leq \mathcal{C}^l_{\vec{AB}} .  \label{CCl}
\end{equation}
The difference $I^L_{\vec{AB}} \equiv \mathcal{C}^l_{\vec{AB}} -\mathcal{C}_{\bar{A}\bar{B}} \geq 0$ is the amount of classical correlation locked in the state $\rho_{AB}$, and it can be unlocked only by a certain amount of classical communication \cite{DHLST04}.

In this article, we are mainly interested in the correlation measures based on optimizations over the most general strategies with POVM measurements. The optimum values may not be achieved by projective measurements; this is illustrated by an example at the end of Section III.

\section{Relations among different measures}

In this section, we shall establish relations for different measures of correlations introduced in the previous section.

For convenience, we define a quantity $\mathcal{S}^m_{AB}$ of a bipartite state $\rho_{AB}$ as
\begin{equation}
\mathcal{S}^m_{AB} (\rho_{AB}) = min\{ S(\rho_A), S(\rho_B), S(A:B) \} .
\end{equation}
We shall show that this minimum is actually an upper bound for various measures of classical correlation in this section.  For two special cases, $\mathcal{S}^m_{AB}$ can be simplified.
If $\rho_{AB}$ is separable, one has $S(\rho_A) \leq S(\rho_{AB})$ and $S(\rho_B) \leq S(\rho_{AB})$ \cite{wu2002}, therefore $\mathcal{S}^m_{AB} = S(A:B)$ for any separable state.
If $\rho_{AB}$ is a pure state, one easily has $\mathcal{S}^m_{AB} = S(\rho_A) =S(\rho_B)$.

In this article, a quantity with an overline denotes the regularized version of the quantity, for example,
\begin{equation}
\overline{\mathcal{C}}_{\bar{A}\bar{B}} (\rho_{AB})\equiv \lim_{n \rightarrow \infty} \frac{1}{n} \mathcal{C}_{\bar{A}\bar{B}} (\rho^{\otimes n}_{AB}) ,
\end{equation}
and similarly for other quantities.
The regularized quantity is defined via local collective measurements, for example, $\mathcal{C}_{\bar{A}\bar{B}} (\rho^{\otimes n}_{AB})$ is obtained by maximization over all local collective measurements, i.e., Alice's measurement could be performed on her $n$ copies of systems together, and Bob's measurement could be performed on his $n$ copies of systems together.
The set of local measurements on individual copies is a subsect of the set of local collective measurements, hence,
$\mathcal{C}_{\bar{A}\bar{B}} (\rho^{\otimes n}_{AB})\geq n \mathcal{C}_{\bar{A}\bar{B}} (\rho_{AB})$. Similar relations hold for other measures of classical correlation introduced in section II.
Therefore, for the measures of classical correlation, we have
\begin{eqnarray}
\overline{\mathcal{C}}_{\bar{A}\bar{B}} \geq  \mathcal{C}_{\bar{A}\bar{B}} \\
\overline{\mathcal{C}}^l_{\vec{AB}} \geq  \mathcal{C}^l_{\vec{AB}} \\
\overline{\mathcal{C}}^h_{\bar{A}B} \geq \mathcal{C}^h_{\bar{A}B} .   \label{ChChreg}
\end{eqnarray}
As either measure of nonclassical correlation, $\mathcal{Q}^s_{AB}$ or $\mathcal{Q}^d_{\bar{A}B}$, is defined as the difference between the quantum mutual information $S(A:B)$ and the corresponding measure of classical correlation $\mathcal{C}_{\bar{A}\bar{B}}$ or $\mathcal{C}^h_{\bar{A}B}$, we have
\begin{eqnarray}
\overline{\mathcal{Q}}^s_{AB} &\leq& \mathcal{Q}^s_{AB} \\
\overline{\mathcal{Q}}^d_{\bar{A}B} &\leq& \mathcal{Q}^d_{\bar{A}B}  .
\end{eqnarray}
Since the von Neumann entropy is additive,  $\mathcal{S}^m_{AB}$ is also additive,
\begin{equation}
\overline{\mathcal{S}}^m_{AB}= \mathcal{S}^m_{AB}  \label{Smadditive}
\end{equation}
for any bipartite state $\rho_{AB}$.

Now we give the following relations for the measures of classical correlation.

{\sl Proposition 1.}  For an arbitrary bipartite state $\rho_{AB}$, we have
\begin{eqnarray}
\mathcal{C}_{\bar{A}\bar{B}} &\leq & \mathcal{C}^l_{\vec{AB}} \leq \mathcal{C}^h_{\bar{A}B} \leq \overline{\mathcal{C}}^h_{\bar{A}B} \leq \mathcal{S}^m_{AB}   \label{th1-1} \\
\overline{\mathcal{C}}_{\bar{A}\bar{B}} &\leq & \overline{\mathcal{C}}^l_{\vec{AB}} \leq \overline{\mathcal{C}}^h_{\bar{A}B} \leq \mathcal{S}^m_{AB} .  \label{th1-2}
\end{eqnarray}
The above relations are still correct if indices $A$ and $B$ are exchanged.
The proof is left to the Appendix.

For the simplest case when $\rho_{AB}$ is a pure state, all the inequalities in proposition 1 become equalities, and all the quantities in (\ref{th1-1}) and (\ref{th1-2}) are equal to the von Neumann entropy of the marginal density matrix on either side.
For general cases, the inequalities in (\ref{th1-1}) establish the order of several measures of classical correlation in a bipartite state $\rho_{AB}$.
The classical correlation $\mathcal{C}_{\bar{A}\bar{B}}$ accessible by local measurements on both A and B is upper bounded by the net gain $\mathcal{C}^l_{\vec{AB}}$ of classical correlation if one-way communication of a classical message from Alice to Bob is allowed in addition to the local measurements. This net gain $\mathcal{C}^l_{\vec{AB}}$ is upper bounded again by the Holevo bound $\mathcal{C}^h_{\bar{A}B}$ of the ensemble prepared for Bob by Alice's local measurement, which is, in turn, upper bounded by its regularized version.  All the measures of classical correlation are upper bounded by $\mathcal{S}^m_{AB}$, which is the minimum of the three quantities $\{ S(\rho_A), S(\rho_B), S(A:B) \}$.
(\ref{th1-2}) gives the order of the corresponding regularized measures.

The following lemma is very useful in the discussions later on.

{\sl Lemma 2.}
Suppose two bipartite states $\rho_{AB}$ and $\sigma_{AB}$ are shared by Alice and Bob, and $\sigma_{AB}$ is separable, then
\begin{eqnarray}
\mathcal{C}^h_{\bar{A}B} (\rho_{AB} \otimes \sigma_{AB}) &=& \mathcal{C}^h_{\bar{A}B} (\rho_{AB} ) +\mathcal{C}^h_{\bar{A}B} (\sigma_{AB})  \label{lemma2-11} \\
\mathcal{C}^h_{A\bar{B}} (\rho_{AB} \otimes \sigma_{AB}) &=& \mathcal{C}^h_{A\bar{B}} (\rho _{AB}) +\mathcal{C}^h_{A\bar{B}} (\sigma_{AB}) , \label{lemma2-22}
\end{eqnarray}
and therefore
\begin{eqnarray}
\mathcal{C}^h_{\bar{A}B}(\sigma_{AB}^{\otimes n}) &=& n \mathcal{C}^h_{\bar{A}B}(\sigma_{AB}) \label{lemma2-33} \\
\mathcal{C}^h_{A\bar{B}}(\sigma_{AB}^{\otimes n}) &=& n \mathcal{C}^h_{A\bar{B}}(\sigma_{AB}) \\
\overline{\mathcal{C}}^h_{\bar{A}B}(\sigma_{AB}) &=&  \mathcal{C}^h_{\bar{A}B}(\sigma_{AB}) \\
\overline{\mathcal{C}}^h_{A\bar{B}}(\sigma_{AB}) &=&  \mathcal{C}^h_{A\bar{B}}(\sigma_{AB}) \label{lemma2-66}
\end{eqnarray}
for any separable state $\sigma_{AB}$.
This lemma was originally proved in \cite{DW08} via inequalities of mutual information, an alternative proof is given in the appendix.

In the rest of this section, we shall study the relations of different measures for a special class of states.

The difference between the symmetric measure $C$ ($\mathcal{Q}^s$) and the asymmetric measure $\mathcal{C}^h$ ($\mathcal{Q}^d$) is illustrated by the classical-quantum (CQ) state
\begin{equation}
\rho^{cq}_{AB} = \sum_i p_i \left|i \right\rangle _A \left\langle i \right| \otimes  \rho_i^B \label{CQdefine}
\end{equation}
where $\left|i \right\rangle _A $ are basis states for system A and $ \rho_i^B$ are arbitrary states of system B.
For this CQ state, one has $\mathcal{S}^m_{AB}(\rho^{cq}_{AB}) = S(A:B) = S(\rho_B) -\sum_i p_i S(\rho_i^B)$, where $\rho_B= \sum_i p_i \rho_i^B$.
From lemma 2, we know that both $\mathcal{C}^h_{\bar{A}B}$ ($\mathcal{Q}^d_{\bar{A}B}$) and $\mathcal{C}^h_{A\bar{B}}$ ($\mathcal{Q}^d_{A\bar{B}}$) are additive for any separable states, hence also for the CQ states, i.e.,
$\mathcal{C}^h_{\bar{A}B}=\overline{\mathcal{C}}^h_{\bar{A}B}$,  $\mathcal{C}^h_{A\bar{B}}=\overline{\mathcal{C}}^h_{A\bar{B}}$,
$\mathcal{Q}^d_{\bar{A}B}=\overline{\mathcal{Q}}^d_{\bar{A}B}$,  $\mathcal{Q}^d_{A\bar{B}}=\overline{\mathcal{Q}}^d_{A\bar{B}}$.
The optimum POVM in the definition of $\mathcal{C}^h_{\bar{A}B}$ is the projective measurement onto $\{\left|i \right\rangle _A \}$, since one can easily show that $\mathcal{C}^h_{\bar{A}B}$ achieves the upper bound $\mathcal{S}^m_{AB}$ via this projective measurement on A.
Therefore, for the CQ state,
\begin{eqnarray}
\mathcal{C}_{\bar{A}\bar{B}} &\leq & \mathcal{C}^l_{\vec{AB}} \leq \mathcal{C}^h_{\bar{A}B} = \overline{\mathcal{C}}^h_{\bar{A}B} =\mathcal{S}^m_{AB} = S(A:B) , \\
\mathcal{Q}^s_{AB} &\geq & \mathcal{Q}^d_{\bar{A}B} =\overline{\mathcal{Q}}^d_{\bar{A}B} =0 ,
\end{eqnarray}
where the inequalities become equalities when the supports of $\{ \rho_i^B\}$ are orthogonal (so $\rho^{cq}_{AB}$ becomes a CC state).
We also have the following proposition for CQ states.

{\sl Proposition 3.}
For the CQ state in (\ref{CQdefine}), we have
\begin{eqnarray}
\mathcal{C}_{\bar{A}\bar{B}}(\rho^{cq}_{AB}) &=& \mathcal{C}^h_{A\bar{B}}(\rho^{cq}_{AB}),  \\
\mathcal{Q}^s_{AB}(\rho^{cq}_{AB}) &=& \mathcal{Q}^d_{A\bar{B}}(\rho^{cq}_{AB}),
\end{eqnarray}
and the optimum POVM measurement that achieves $\mathcal{C}_{\bar{A}\bar{B}}(\rho^{cq}_{AB})$ actually involves a local projective measurement (by Alice) onto the basis states $\left|i \right\rangle_A$ and a local POVM measurement (by Bob) which is the optimum POVM measurement to achieve $\mathcal{C}^h_{A\bar{B}}(\rho^{cq}_{AB})$; furthermore, $\mathcal{C}_{\bar{A}\bar{B}}$ and $\mathcal{Q}^s_{AB}$ are also additive for the CQ states,
\begin{eqnarray}
\overline{\mathcal{C}}_{\bar{A}\bar{B}}(\rho^{cq}_{AB}) &=& \mathcal{C}_{\bar{A}\bar{B}}(\rho^{cq}_{AB}),  \\
\overline{\mathcal{Q}}^s_{AB}(\rho^{cq}_{AB}) &=& \mathcal{Q}^s_{AB}(\rho^{cq}_{AB})  .
\end{eqnarray}
Proof is left to the appendix.

As a summary of the results for the CQ state $\rho^{cq}_{AB}$ in (\ref{CQdefine}), we have
\begin{eqnarray}
&&\overline{\mathcal{C}}_{\bar{A}\bar{B}} =\mathcal{C}_{\bar{A}\bar{B}} = \mathcal{C}^h_{A\bar{B}} =\overline{\mathcal{C}}^h_{A\bar{B}} \nonumber \\
&\leq & \mathcal{C}^h_{\bar{A}B} = \overline{\mathcal{C}}^h_{\bar{A}B} =\mathcal{S}^m_{AB} = S(A:B)
\end{eqnarray}
and
\begin{equation}
\overline{\mathcal{Q}}^s_{AB}= \mathcal{Q}^s_{AB} = \mathcal{Q}^d_{A\bar{B}} =\overline{\mathcal{Q}}^d_{A\bar{B}} \geq  \mathcal{Q}^d_{\bar{A}B} =\overline{\mathcal{Q}}^d_{\bar{A}B} =0 .
\end{equation}
In general, the inequalities could be strict when some supports of $\{ \rho_i^B\}$ in (\ref{CQdefine}) are not orthogonal.

Before leaving this section, we point out that strategies based on projective measurements may not be able to extract all the classical correlation in a state, general POVM measurements may indeed have advantages. As an example, we consider a particular CQ state
\begin{equation}
\rho_{AB}^{cq} =\sum_{i=1}^3 \frac{1}{3} \left|i \right\rangle_A \left\langle i \right| \otimes \left| \phi _i \right\rangle _B \left\langle \phi_i \right|  \label{povmvspvm}
\end{equation}
where system A is a qutrit with the basis states $\left| i \right\rangle_A$ and system B is a qubit with the pure states $\left| \phi _i \right\rangle _B$ forming equal angles $\frac{2\pi}{3}$ in the same plane according to a Bloch sphere picture. It is shown in \cite{wu2009correl} that $\mathcal{C}_{\bar{A}\bar{B}}$ (a different symbol $I_{max}$ is used in \cite{wu2009correl}) is strictly greater than the corresponding measure based on projective measurements only. From proposition 3, we immediately know that $\mathcal{C}^h_{A\bar{B}}$ cannot be achieved by projective measurements on system B either. Therefore, for the state in (\ref{povmvspvm}), strategies based on projective measurements are not enough to achieve $\mathcal{C}_{\bar{A}\bar{B}}$, $\mathcal{C}^h_{A\bar{B}}$, $\mathcal{Q}^s_{AB}$, and $ \mathcal{Q}^d_{A\bar{B}}$, POVM measurements are indeed necessary!  However, it is easy to see that each measure of classical (quantum) correlation based on projective measurements provides a lower (upper) bound of the corresponding measure based on general POVM measurements.

\section{Entanglement irreversibility}

In section III, we have derived relations among different measures of classical correlation and nonclassical correlation beyond entanglement.
In this section, we discuss a somehow different but related problem,  i.e., entanglement irreversibility, with the help of the relations we have derived.

For a bipartite pure state $\left| \psi \right\rangle_{AB}$, entanglement of formation $E^F$ is simply the von Neumann entropy of the marginal density matrix on either side.  A general bipartite state $\rho_{AB}$ can be decomposed into an ensemble of bipartite pure states $\rho_{AB} = \sum_k q_k \left|\psi_k \right\rangle_{AB} \left\langle \psi_k \right| \}$, this decomposition is not unique in general (except when $\rho_{AB}$ itself is a pure state). Entanglement of formation $E^F(\rho_{AB})$ is defined as the minimal average pure-state entanglement over all possible decompositions of $\rho_{AB}$.
Entanglement cost $E^C(\rho_{AB})$ denotes the number of singlets needed to generate $\rho_{AB}$ per copy via LOCC in the process of entanglement dilution, it is equal to the regularized version of entanglement of formation \cite{HHT01}, i.e., $E^C (\rho_{AB}) = \overline{E}^F (\rho_{AB}) \equiv \lim_{n \rightarrow \infty} \frac{1}{n} E^F (\rho^{\otimes n}_{AB})$.  Entanglement of distillation $E^D(\rho_{AB})$ denotes the number of singlets that can be generated asymptotically per copy of $\rho_{AB}$ via LOCC.

Entanglement dilution and entanglement distillation are generally irreversible, i.e., $E^C \geq E^D$ with strict inequality for many cases.
In the following, we shall discuss how entanglement irreversibility is related to the measures of classical and nonclassical correlations we have discussed in the previous sections.

For any tripartite pure state $\left| \Psi \right\rangle_{ABC}$, the entanglement of formation $E^F$ in $\rho_{AB}$ and a measure of classical correlation $\mathcal{C}^h_{A\bar{C}}$ in $\rho_{AC}$ have the Koashi-Winter relation \cite{KW04}
\begin{equation}
E^F(\rho_{AB})+\mathcal{C}^h_{A\bar{C}}(\rho_{AC}) =  S(\rho_A) .  \label{EFCrelationpure}
\end{equation}
In order to get the regularized version of this equality, we consider $n$ copies of the state $\left| \Psi \right\rangle_{ABC}$.  The total state is still a tripartite pure state, therefore, $E^F(\rho_{AB}^{\otimes n})+\mathcal{C}^h_{A\bar{C}}(\rho_{AC}^{\otimes n}) =  S(\rho_A^{\otimes n})=n S(\rho_A)$. Considering the equality $\frac{1}{n} E^F(\rho_{AB}^{\otimes n})+\frac{1}{n}  \mathcal{C}^h_{A\bar{C}}(\rho_{AC}^{\otimes n}) = S(\rho_A)$ in the limit $n \rightarrow \infty$, we immediately obtain the regularized version
\begin{equation}
E^C(\rho_{AB})+\overline{\mathcal{C}}^h_{A\bar{C}}(\rho_{AC}) =  S(\rho_A)  . \label{EFCrelationregularized}
\end{equation}
For convenience, the coherent information of $\rho_{AB}$ is defined as
\begin{equation}
I^C(\rho_{AB}) = max\{ 0, S(\rho_A)-S(\rho_{AB}), S(\rho_B)-S(\rho_{AB}) \} . \label{cohinfdef}
\end{equation}
The coherent information is a lower bound of entanglement of distillation $E^D(\rho_{AB})$ \cite{Bennett96hash,DW04},
\begin{equation}
E^D(\rho_{AB}) \geq  I^C(\rho_{AB}) .  \label{EDIrelation}
\end{equation}

Now we present the following proposition that gives an upper bound for entanglement irreversibility as well as a sufficient condition for reversible entanglement.

{\sl Proposition 4.}  Let $\rho_{AB}$, $\rho_{AC}$, and $\rho_{BC}$ be the three reduced bipartite states of a tripartite pure state $\left|\Psi \right\rangle_{ABC}$;  then \\
(i) entanglement cost satisfies the equality
\begin{equation}
E^C(\rho_{AB}) - I^C(\rho_{AB}) =  \mathcal{S}^m(\rho_{AC}) - \overline{\mathcal{C}}^h_{A\bar{C}}(\rho_{AC});  \label{ent1}
\end{equation}
(ii) entanglement irreversibility has an upper bound
\begin{equation}
E^C(\rho_{AB}) -E^D(\rho_{AB}) \leq   \mathcal{S}^m(\rho_{AC}) - \overline{\mathcal{C}}^h_{A\bar{C}}(\rho_{AC});  \label{ent2}
\end{equation}   \\
(iii) entanglement in $\rho_{AB}$ is reversible, i.e., $E^D(\rho_{AB}) =E^C(\rho_{AB})$, if the regularized measure of classical correlation $\overline{\mathcal{C}}^h_{A\bar{C}}$ in $\rho_{AC}$ reaches its upper bound $\mathcal{S}^m_{AC}$, i.e., $\overline{\mathcal{C}}^h_{A\bar{C}} = \mathcal{S}^m_{AC}$;  \\
(iv) the above three statements are still valid when each subscript $AC$ is replaced with $BC$.

Proof is left to the appendix.

\begin{figure}
\centering
\includegraphics[width=8cm]{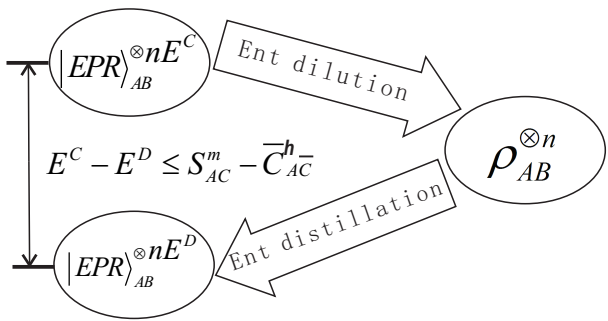}
\caption{Entanglement irreversibility $E^C(\rho_{AB})-E^D(\rho_{AB})$ has an upper bound $\mathcal{S}^m(\rho_{AC})-\overline{\mathcal{C}}^h_{A\bar{C}}(\rho_{AC})$, where $\rho_{AC}$ is an arbitrary state that has the same purification $\left|\Psi \right\rangle_{ABC}$ as $\rho_{AB}$.}
\label{fig2}
\end{figure}

Eqs. (\ref{ent1}) and (\ref{ent2}) provide some insights into entanglement irreversibility (see Fig. 1), the entanglement irreversibility of $\rho_{AB}$ is bounded from above by the discrepancy between the regularized measure $\overline{\mathcal{C}}^h_{A\bar{C}}$ of classical correlation and its upper bound $\mathcal{S}^m$ in $\rho_{AC}$, where $\rho_{AC}$ is the reduced state that shares with $\rho_{AB}$ the same tripartite purification $\left|\Psi \right\rangle_{ABC}$.   Although proposition 4 does not tell us how to discriminate all reversible entanglement from irreversible entanglement, it does provide a sufficient condition for reversible entanglement.

It is suggested in \cite{COF11} that entanglement irreversibility of $\rho_{AB}$ is due to nonzero regularized quantum discord $\overline{\mathcal{Q}}^h_{A\bar{C}}$.
From (\ref{ent2}) we have
\begin{equation}
E^C(\rho_{AB}) -E^D(\rho_{AB}) \leq   S(A:C) - \overline{\mathcal{C}}^h_{A\bar{C}} = \overline{\mathcal{Q}}^d_{A\bar{C}} .  \label{entqd}
\end{equation}
Therefore, the regularized quantum discord $\overline{\mathcal{Q}}^d_{A\bar{C}}(\rho_{AC})$ is indeed an upper bound of $E^C(\rho_{AB}) -E^D(\rho_{AB})$. However, we point out that it is only an upper bound by the following example.
We consider a tripartite pure state $\Psi_{ABC}$ (shared among Alice, Bob and Camilla), which is constructed in the following way: Alice, Bob and Camilla share a Greenberger-Horne-Zeilinger (GHZ) state together, in addition, Alice and Bob share an Einstein-Podolski-Rosen (EPR) state, while Alice and Camilla share another EPR state, i.e.,
\begin{equation}
\left| \Psi \right\rangle _{ABC} =\left| GHZ\right\rangle_{a_1 b_1 c_1} \left| EPR \right\rangle_{a_2 b_2} \left| EPR \right\rangle_{a_3 c_2} .
\end{equation}
Here $\left| EPR \right\rangle =\frac{1}{\sqrt{2}} ( \left| 00 \right\rangle + \left| 11 \right\rangle )$ and $\left| GHZ\right\rangle =\frac{1}{\sqrt{2}} ( \left| 000 \right\rangle + \left| 111 \right\rangle )$.
It is straightforward to obtain $\mathcal{S}^m_{AC} =2 $. One also has $\mathcal{C}^h_{A\bar{C}}\geq 2$ since $2$ can be reached by a particular choice of Camilla's local measurement: projecting $c_1$ onto the Schmidt basis of the GHZ state, and $c_2$ onto an arbitrary basis.
From proposition 1 we know that $ \mathcal{C}^h_{A\bar{C}} \leq \overline{\mathcal{C}}^h_{A\bar{C}} \leq \mathcal{S}^m_{AC}$, together with $\mathcal{C}^h_{A\bar{C}}\geq 2$ and $\mathcal{S}^m_{AC} =2 $, we obtain $\mathcal{C}^h_{A\bar{C}}=\overline{\mathcal{C}}^h_{A\bar{C}} =\mathcal{S}^m_{AC} =2 $. Therefore, entanglement in $\rho_{AB}$ is reversible according to proposition 4. Entanglement reversibility can also be directly verified
\begin{equation}
E^C(\rho_{AB}) = E^D (\rho_{AB}) =1
\end{equation}
as we can easily distillate one copy of EPR from one copy of $\rho_{AB}$ via LOCC, and create one copy of $\rho_{AB}$ from one copy of EPR via LOCC as well.
However, the regularized quantum discord
\begin{equation}
\overline{\mathcal{Q}}^d_{A\bar{C}} = S(A:C) - \overline{\mathcal{C}}^h_{A\bar{C}} =3-2 =1
\end{equation}
does not vanish! Therefore, the regularized quantum discord $\overline{\mathcal{Q}}^d_{A\bar{C}}$ is just an upper bound for entanglement irreversibility of $\rho_{AB}$.

One may further attempt to ask whether the difference $\mathcal{S}^m(\rho_{AC}) - \overline{\mathcal{C}}^h_{A\bar{C}}(\rho_{AC})$ characterizes entanglement irreversibility $E^C(\rho_{AB}) -E^D(\rho_{AB})$, instead of being just its upper bound. We consider another example: Suppose Alice, Bob and Camilla share the same state as in the previous example, in addition, Bob and Camilla share another EPR state as well, i.e., the overall state is
\begin{equation}
\left| \Phi \right\rangle _{ABC} =\left| GHZ\right\rangle_{a_1 b_1 c_1} \left| EPR \right\rangle_{a_2 b_2} \left| EPR \right\rangle_{a_3 c_2} \left| EPR \right\rangle_{b_3 c_3} .  \label{triplephi}
\end{equation}
We show that
\begin{eqnarray}
E^C(\rho_{AB}) &=&  E^D(\rho_{AB}) =1   \label{PhiECED}    \\
\mathcal{S}^m(\rho_{AC}) &=3&   \label{PhiSm} \\
\overline{\mathcal{C}}^h_{A\bar{C}}(\rho_{AC}) &=& 2  \label{PhiChregAC}
\end{eqnarray}
in the appendix.
In this case, one can easily have
\begin{equation}
0= E^C(\rho_{AB}) -E^D(\rho_{AB}) <  \mathcal{S}^m(\rho_{AC}) - \overline{\mathcal{C}}^h_{A\bar{C}}(\rho_{AC})=1 .
\end{equation}
Therefore, $\mathcal{S}^m(\rho_{AC}) - \overline{\mathcal{C}}^h_{A\bar{C}}(\rho_{AC})$ is just an upper bound of entanglement irreversibility of $\rho_{AB}$ as well.

\section{Additivity}

The quantity $\mathcal{S}^m$ is additive but not fully additive, i.e, $\mathcal{S}^m(\rho^{\otimes n}_{AB}) = n \mathcal{S}^m(\rho_{AB})$ while
$\mathcal{S}^m(\rho_{AB} \otimes \sigma_{AB}) \neq  \mathcal{S}^m(\rho_{AB})+\mathcal{S}^m(\sigma_{AB})$.
In fact, we have
\begin{equation}
\mathcal{S}^m(\rho_{AB} \otimes \sigma_{AB}) \geq \mathcal{S}^m(\rho_{AB})+\mathcal{S}^m(\sigma_{AB})  \label{Smnotfulladditive}
\end{equation}
for arbitrary bipartite states $\rho_{AB}$ and $\sigma_{AB}$. The inequality looks surprising since $\mathcal{S}^m$ is the minimum of three quantities, each of which is fully additive as the von Neumann entropy is fully additive. However, it becomes obvious if one notices the following fact,
\begin{eqnarray}
min \{ S(\rho_A)+ S(\sigma_A), S(\rho_B)+ S(\sigma_B), \nonumber \\
S(A:B)|_\rho + S(A:B)|_\sigma \} \nonumber \\
\geq min \{ S(\rho_A), S(\rho_B), S(A:B)|_\rho \} + \nonumber \\
min \{  S(\sigma_A), S(\sigma_B), S(A:B)|_\sigma \}
\end{eqnarray}
since the minimum values may be reached at different places.  One has an example of strict inequality if one considers the marginal state of (\ref{triplephi}) by tracing out C.

In general, the measures for classical correlation ($C$, $\mathcal{C}^l$, $\mathcal{C}^h$), and the measures for nonclassical correlation ($\mathcal{Q}^s$, $\mathcal{Q}^d$) may not be additive. Their regularized versions are additive by definition, but there is no indication that the regularized versions are fully additive.

\begin{figure}
\centering
\includegraphics[width=6cm]{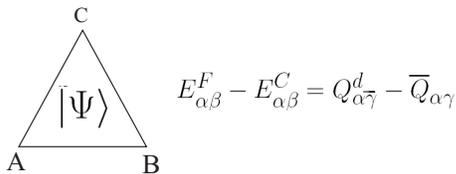}
\caption{For a tripartite pure state $\left| \Psi \right\rangle_{ABC}$,  the additivity of entanglement of formation in $\rho_{\alpha \beta}$ is equivalent to the additivity of quantum discord in another bipartite state $\rho_{\alpha \gamma}$ (or $\rho_{\beta \gamma}$) when the measurement is performed on the third system $\gamma$.}
\label{fig3}
\end{figure}

There is an equivalence for the additivity of the three quantities: entanglement of formation $E^F$, the measure of classical correlation $\mathcal{C}^h$, and quantum discord $\mathcal{Q}^d$.  For any tripartite pure state $\left| \Psi \right\rangle_{ABC}$, from (\ref{EFCrelationpure}), (\ref{EFCrelationregularized}) and the definition of quantum discord, we have
\begin{eqnarray}
E^F(\rho_{AB}) - E^C(\rho_{AB}) &=& \overline{\mathcal{C}}^h_{A\bar{C}}(\rho_{AC}) - \mathcal{C}^h_{A\bar{C}}(\rho_{AC}) \nonumber  \\
&=& \mathcal{Q}^d_{A\bar{C}} (\rho_{AC}) -\overline{\mathcal{Q}}^d_{A\bar{C}}(\rho_{AC}) \nonumber  \\
&=& \overline{\mathcal{C}}^h_{B\bar{C}}(\rho_{BC}) - \mathcal{C}^h_{B\bar{C}}(\rho_{BC})  \nonumber  \\
&=& \mathcal{Q}^d_{B\bar{C}} (\rho_{BC}) -\overline{\mathcal{Q}}^d_{B\bar{C}}(\rho_{BC})  .  \;  \label{EFnonadditiverelations}
\end{eqnarray}
From these relations we know that the nonadditivity of entanglement of formation for a bipartite state $\rho_{AB}$ is equivalent to the nonadditivity of quantum discord $\mathcal{Q}^d$ (or the classical correlation $\mathcal{C}^h$) in either $\rho_{AC}$ or $\rho_{BC}$ as long as they share the same purification $\left| \Psi \right\rangle_{ABC}$ and the measurement is performed on the third system C (see Fig. 2).  In this way, if one finds a bipartite state with $\mathcal{C}^h\neq \overline C^h$ (or $\mathcal{Q}^d \neq \overline{\mathcal{Q}}^d$), he finds a corresponding bipartite state with nonadditive entanglement of formation, and vice versa.

For the CQ state in (\ref{CQdefine}), a purification of $\rho_{AB}$ is $\left| \Psi \right\rangle_{ABC} =\sum_i \sqrt{p_i} \left| i \right\rangle_A \otimes \left| \psi_i \right\rangle_{BC_1} \otimes \left| i\right\rangle_{C_2}$, where both subsystems $C_1$ and $C_2$ are held by Camilla, and each $\left| \psi_i \right\rangle_{B\mathcal{C}_1}$ is a purification of the corresponding $\rho_i^B$, i.e., $\rho_i^B = tr_{C_1} (\left| \psi_i \right\rangle_{BC_1} \left\langle \psi_i \right|)$. The other two reduced states are given by
\begin{eqnarray}
\rho_{AC} &=& \sum_{ij} \sqrt{p_i p_j} \left|i \right\rangle _A \left\langle j \right| \otimes tr_B(\left| \psi_i \right\rangle_{BC_1} \left\langle \psi_j \right|) \otimes \left| i \right\rangle_{C_2} \left\langle j \right|  \nonumber \\
\rho_{BC} &=& \sum_{i} p_i \left| \psi_i \right\rangle_{BC_1} \left\langle \psi_i \right| \otimes \left|i \right\rangle _{C_2} \left\langle i \right| \nonumber
\end{eqnarray}
where $\rho_{BC}$ is called a pseudopure state \cite{horodecki09rev}.
From lemma 2, one has $\mathcal{C}^h_{\bar{A}B} = \overline{\mathcal{C}}^h_{\bar{A}B}$ and $\mathcal{C}^h_{A\bar{B}} = \overline{\mathcal{C}}^h_{A\bar{B}}$. Therefore, entanglement of formation in both $\rho_{BC}$ and $\rho_{AC}$ are additive, i.e., $E^F(\rho_{BC})= E^C(\rho_{BC})$ and $E^F(\rho_{AC})= E^C(\rho_{AC})$.

As another example, consider the separable state
\begin{equation}
\rho_{AB} = \sum_i p_i \left| a_i \right\rangle \left\langle a_i \right| \otimes \left| b_i \right\rangle \left\langle b_i \right|
\end{equation}
where $\{ \left| a_i \right\rangle \}$ and $\{ \left| b_i \right\rangle \}$ are normalized (nonorthogonal in general) states of A and C.
According to lemma 2, the measure of classical correlation $\mathcal{C}^h_{\bar{A}B}$ ($\mathcal{C}^h_{A\bar{B}}$) and quantum discord $\mathcal{Q}^d_{\bar{A}B}$ ($\mathcal{Q}^d_{A\bar{B}}$) are additive for $\rho_{AB}$. Construct the purification $\left| \psi \right\rangle_{ABC} = \sum_i \sqrt{p_i} \left| a_i \right\rangle_A \left| b_i \right\rangle_B \left| i \right\rangle_C$, and we immediately know that the entanglement of formation is additive for $\rho_{AC}$, i.e., $E^C(\rho_{AC})=E^F(\rho_{AC})$ with
\begin{equation}
\rho_{AC} = \sum_{ij} \sqrt{p_i p_j} \left\langle b_j | b_i \right\rangle \left| a_i \right\rangle_A \left\langle a_j \right| \otimes \left| i\right\rangle_C \left\langle j \right|  .  \label{owmcs}
\end{equation}
The state in (\ref{owmcs}) is called the one-way maximally correlated state in \cite{COF11}.

\section{Conclusion}

In conclusion, we have studied the properties of several naturally defined measures of correlations and provided insights into some open questions. We have obtained inequality relations among several different measures of classical and nonclassical correlations as well as equivalence relation of different measures for certain states. We consider the measures that are defined according to optimizations over POVM measurements, they are different from the corresponding measures based on projective measurements in general.
We have derived a sufficient condition for reversible entanglement as well as an upper bound for entanglement irreversibility. We have also discussed the additivity relations of entanglement of formation, quantum discord and a certain measure of classical correlation.  We hope that the results and discussions here could provide useful insights into the open problems in quantum information theory.

\section*{Acknowledgements}

The author would like to thank B. Sanders for his hospitality during the author's visit to IQIS, M. C. de Oliveira for drawing his attention to ref. \cite{COF11}, R. B. Griffiths, D. Yang, L. Yu, and P. J. Coles for valuable discussions.  The author acknowledges support from a USTC visiting funding and AITF for a visiting professorship.
This research also receives partial support from NSFC (Grant No. 11075148).

\section*{Appendix}
\subsection{Proof of proposition 1}

It is sufficient to show (\ref{th1-1}), since the regularized version (\ref{th1-2}) follows directly from (\ref{th1-1}), the additivity of $\mathcal{S}^m$ and the definition of the regularized quantities.
Let us consider the inequalities in (\ref{th1-1}) one by one. The first inequality $\mathcal{C}_{\bar{A}\bar{B}} \leq \mathcal{C}^l_{\vec{AB}}$ is shown in (\ref{CCl}) as $\mathcal{C}_{\bar{A}\bar{B}}$ can be viewed as $\mathcal{C}^l_{\vec{AB}}$ without communication. The second inequality needs to be proved below.
The third inequality $\mathcal{C}^h_{\bar{A}B} \leq \overline{\mathcal{C}}^h_{\bar{A}B}$ is shown in (\ref{ChChreg}).
Since $\mathcal{S}^m_{AB}$ is additive, the fourth inequality $\overline{\mathcal{C}}^h_{\bar{A}B} \leq \mathcal{S}^m_{AB}$ follows from $\mathcal{C}^h_{\bar{A}B} \leq \mathcal{S}^m_{AB}$ (that needs to be proved below) via the regularization. In other words, we only need to prove the following two relations:
\begin{eqnarray}
\mathcal{C}^l_{\vec{AB}} \leq \mathcal{C}^h_{\bar{A}B}  \label{appA-1}\\
\mathcal{C}^h_{\bar{A}B} \leq \mathcal{S}^m_{AB} .   \label{appA-2}
\end{eqnarray}
For a bipartite state $\rho_{AB}$, $\mathcal{C}^l_{\vec{AB}} (\rho_{AB}) = \max_{ \mathcal{E}} [I(A:B)(\mathcal{E})-ccc(\mathcal{E})]$, and $\mathcal{C}^h_{\bar{A}B} (\rho_{AB}) = \max_{ \{ \Pi _j^{A} \} } [ S(\rho_B) - \sum_j p_j S(\rho_j^B) ]$.

We first prove (\ref{appA-1}).
Without loss of generality, suppose the optimum LOCC protocol $\mathcal{E}_{opt}$ that achieves the maximum in the definition of $\mathcal{C}^l_{\vec{AB}}$ does the following. First, Alice performs a local POVM measurement $\{ \Pi_i^A  \}$ on system A.  Whenever Alice gets the $i$-th outcome, which occurs with probability $p_i= tr_{AB} ((\Pi_i^A \otimes I_B)\rho_{AB})$, system B is left in the state $\rho_i^B = tr_{A} ((\Pi_i^A \otimes I_B)\rho_{AB})/p_i$. The one-way classical communication from Alice to Bob, which can always be carried out after Alice's measurement, could depend on Alice's measurement result $i$. A classical message of finite number of bits could be modeled as an integer function $f(i)$ of Alice's measurement result $i$. We construct the following tripartite state:
\begin{equation}
\sigma_{ABC} =\sum_i p_i \left|i \right\rangle _A \left\langle i \right| \otimes \left| f(i) \right\rangle _C \left\langle f(i) \right| \otimes \rho_i^B .
\end{equation}
Therefore, for this optimal LOCC protocol $\mathcal{E}_{opt}$,
\begin{eqnarray}
I(A:B)(\mathcal{E}_{opt}) &\leq & S(A:CB)(\sigma_{ABC}) \nonumber \\
&=& S(\sigma_A) +S(\sigma_{CB}) -S(\sigma_{ACB}) \nonumber \\
&=& S(\sigma_{AC}) +S(\sigma_{CB}) -S(\sigma_{ACB})
\end{eqnarray}
where the last equality is due to the fact $S(\sigma_A)=S(\sigma_{AC})$. As $S(\sigma_{CB}) \leq S(\sigma_{C}) +S(\sigma_{B})$, we have
\begin{eqnarray}
I(A:B)(\mathcal{E}_{opt}) &\leq & S(\sigma_{AC}) +S(\sigma_{C}) +S(\sigma_{B}) -S(\sigma_{ACB}) \nonumber \\
&=& S(AC:B) + S(\sigma_C) \nonumber \\
&=& S(\rho_B) -\sum_i p_i S(\rho_i^B) +S(\sigma_C).
\end{eqnarray}
Here, $\rho_B =\sum_i p_i \rho_i^B = \sigma_B$, and $\sigma_C = \sum_i p_i \left| f(i) \right\rangle _C \left\langle f(i) \right|$. $S(\sigma_C)$ denotes the number of classical bits that need to be sent from Alice to Bob in the asymptotic limit. For a single copy, $ccc \geq S(\sigma_C)$, therefore,
\begin{eqnarray}
\mathcal{C}^l_{\vec{AB}} (\rho_{AB}) &=& I(A:B)(\mathcal{E}_{opt})-ccc(\mathcal{E}_{opt}) \nonumber \\
&\leq & I(A:B)(\mathcal{E}_{opt}) - S(\sigma_C) \nonumber \\
&=& S(\rho_B) -\sum_i p_i S(\rho_i^B) \nonumber \\
&\leq & \max_{\{\Pi_j^{\prime A}\}} [ S(\rho_B) -\sum_j p_j^\prime S(\rho_j^{\prime B}) ] \nonumber \\
&=& \mathcal{C}^h_{\bar{A}B}.
\end{eqnarray}
This completes the proof of (\ref{appA-1}).

Next, we prove (\ref{appA-2}), i.e., $\mathcal{C}^h_{\bar{A}B} \leq \mathcal{S}^m_{AB}=min \{ S(\rho_A), S(\rho_B), S(A:B) \}$.
One easily has $\mathcal{C}^h_{\bar{A}B} \leq  S(\rho_B)$ from the definition of $\mathcal{C}^h_{\bar{A}B}$, and one has $\mathcal{C}^h_{\bar{A}B} \leq  S(A:B)$ since quantum discord is non-negative \cite{OZ01}. It is not so straightforward to get $\mathcal{C}^h_{\bar{A}B} \leq  S(\rho_A)$. However, the overall inequality (\ref{appA-2}) can also be proved as follows. Considering a purification of $\rho_{AB}$, from (\ref{EFCrelationpure}) one has
\begin{equation}
\mathcal{C}^h_{\bar{A}B} =  S(\rho_B)- E^F(\rho_{BC}) .  \label{ChSbEf}
\end{equation}
The definition of the coherent information of $\rho_{BC}$ is similar to (\ref{cohinfdef}):
\begin{eqnarray}
I^C(\rho_{BC}) &=& max\{ 0, S(\rho_B)-S(\rho_{BC}), S(\rho_C)-S(\rho_{BC}) \}  \nonumber \\
&=& S(\rho_B)- \mathcal{S}^m_{AB} . \label{cohinfdef11BC}
\end{eqnarray}
On the other hand, the inequality corresponding to (\ref{EDIrelation}) reads as
\begin{equation}
E^D(\rho_{BC}) \geq  I^C(\rho_{BC}) .  \label{EDIrelation11BC}
\end{equation}
Hence,
\begin{equation}
E^F(\rho_{BC}) \geq E^D(\rho_{BC}) \geq I^C(\rho_{BC})= S(\rho_B)- \mathcal{S}^m_{AB} . \label{EfEdIcSb}
\end{equation}
From (\ref{ChSbEf}) and (\ref{EfEdIcSb}), one immediately gets (\ref{appA-2}).

Thus, the proof of proposition 1 is completed.

\subsection{Proof of lemma 2}

We first give the following proposition.

{\sl Proposition 5.}
For an arbitrary ensemble of product states $\{ \rho_k \otimes \sigma_k \}$ with the corresponding probability distribution $\{ p_k\}$,
\begin{equation}
\sum_k p_k S(\rho_k) +S(\sum_k p_k \sigma_k)  \leq S(\sum_k p_k \rho_k \otimes \sigma_k) .  \label{lemma2eq}
\end{equation}
This inequality is a property of the von Neumann entropy, it could actually serve as a separability criterion, which is interesting by itself. The proof is straightforward.
One directly obtains (\ref{lemma2eq}) from the strong subadditivity relation $S(\rho_{ABC})+S(\rho_B) \leq S(\rho_{AB}) +S(\rho_{BC})$ on the tripartite state $\sum_k p_k \rho_k^A \otimes \sigma_k^B \otimes \left| k \right\rangle^C\left\langle k \right|$.

Next, we prove lemma 2.
Alice and Bob share the state $\rho_{A_1 B_1} \otimes \sigma_{A_2 B_2}$, and $\sigma_{A_2 B_2}$ is separable and can always be written as $\sigma_{A_2 B_2} = \sum_k q_k \sigma_k^{A_2} \otimes \sigma_k^{B_2}$.
Suppose $\{ \Pi_i^{*A_1A_2}  \}$ is the optimum POVM measurement performed on $A_1 A_2$ in the definition of $\mathcal{C}^h_{\bar{A}B}$.  Let $p_{i|k}= tr_{A_1 A_2 B_1} (\Pi_i^{*A_1 A_2} \rho_{A_1B_1} \otimes \sigma_k^{A_2})$, and $\rho_{i|k}^{B_1}= tr_{A_1 A_2} (\Pi_i^{*A_1 A_2} \rho_{A_1B_1} \otimes \sigma_k^{A_2})/p_{i|k}$.
When Alice obtains the $i$th result, which occurs with probability $p_i =\sum_k q_k p_{i|k}$, the state of Bob's combined system $B_1 B_2$ is left in the state
$\rho_i^{B_1 B_2} = \sum_k p_{k|i} \rho_{i|k}^{B_1} \otimes \sigma_k^{B_2}$ where $p_{k|i} = q_k p_{i|k}/ p_i$.

We have
\begin{eqnarray}
& & \mathcal{C}^h_{\bar{A}B} (\rho_{A_1 B_1} \otimes \sigma_{A_2 B_2}) \nonumber \\
&=& S(\rho_{B_1}\otimes \sigma_{B_2}) - \sum_i p_i S(\rho_i^{B_1B_2})  \nonumber \\
&=& S(\rho_{B_1})+ S(\sigma_{B_2}) - \sum_i p_i S(\sum_k p_{k|i} \rho_{i|k}^{B_1} \otimes \sigma_k^{B_2}) \nonumber \\
& \leq & S(\rho_{B_1})+ S(\sigma_{B_2}) -  \nonumber \\
&& \sum_i p_i \{ \sum_k p_{k|i} S(\rho_{i|k}^{B_1}) + S(\sum_k p_{k|i} \sigma_k^{B_2})\}   \nonumber \\
&=&  S(\rho_{B_1})- \sum_{ki} q_k p_{i|k} S(\rho_{i|k}^{B_1}) \nonumber \\
&& + S(\sigma_{B_2}) - \sum_i p_i S(\sum_k p_{k|i} \sigma_k^{B_2})      \nonumber \\
&=& \sum_k q_k \{ S(\rho_{B_1})- \sum_{i} p_{i|k} S(\rho_{i|k}^{B_1}) \} \nonumber \\
& & + S(\sigma_{B_2}) - \sum_i p_i S(\sum_k p_{k|i} \sigma_k^{B_2})
\end{eqnarray}
where the inequality is due to proposition 5.
$p_{i|k}$ and $\rho_{i|k}^{B_1}$ can be viewed as the probability and resulting state of obtaining the $i$th result in a measurement on subsystem $A_1$ alone (with subsystem $A_2$ prepared in the state $\sigma_k^{A_2}$ as an ancilla), therefore,
$S(\rho_{B_1})- \sum_{i} p_{i|k} S(\rho_{i|k}^{B_1}) \leq \mathcal{C}^h_{\bar{A}B}(\rho_{A_1 B_1})$.
Let $ \sum_k p_{k|i} \sigma_k^{B_2}= \sigma_i^{B_2}$, and one has
\begin{eqnarray}
\sigma_i^{B_2} &=& \frac{1}{p_i} \sum_k q_k tr_{A_1 A_2 B_1} (\Pi_i^{*A_1 A_2} \rho_{A_1B_1} \otimes \sigma_k^{A_2})   \sigma_k^{B_2}  \nonumber \\
&=& \frac{1}{p_i} \sum_k q_k tr_{A_1 A_2 } (\Pi_i^{*A_1 A_2} \rho_{A_1} \otimes \sigma_{A_2 B_2})
\end{eqnarray}
with
\begin{eqnarray}
p_i &=&  \sum_k q_k tr_{A_1 A_2 B_2} (\Pi_i^{*A_1 A_2} \rho_{A_1} \otimes \sigma_{A_2 B_2}) \nonumber \\
&=&  \sum_k q_k tr_{A_1 A_2} (\Pi_i^{*A_1 A_2} \rho_{A_1} \otimes \sigma_{A_2}) .
\end{eqnarray}
Therefore, $p_i$ and $\sigma_i^{B_2}$ could be viewed as the probability and resulting state of $B_2$ when the $i$th result is obtained in a measurement on subsystem $A_2$ alone with $A_1$ prepared in $\rho_{A_1}$ as an acilla. We have
$S(\sigma_{B_2}) - \sum_i p_i S(\sigma_i^{B_2}) \leq \mathcal{C}^h_{\bar{A}B} (\sigma_{A_2 B_2})$.  By combining the above results, we have
\begin{equation}
\mathcal{C}^h_{\bar{A}B} (\rho_{A_1 B_1} \otimes \sigma_{A_2 B_2}) \leq \mathcal{C}^h_{\bar{A}B} (\rho_{A_1 B_1}) +\mathcal{C}^h_{\bar{A}B} (\sigma_{A_2 B_2})
\end{equation}
which together with the obvious relation
\begin{equation}
\mathcal{C}^h_{\bar{A}B} (\rho_{A_1 B_1} \otimes \sigma_{A_2 B_2}) \geq  \mathcal{C}^h_{\bar{A}B} (\rho_{A_1 B_1}) +\mathcal{C}^h_{\bar{A}B} (\sigma_{A_2 B_2})
\end{equation}
implies that
\begin{equation}
\mathcal{C}^h_{\bar{A}B} (\rho_{A_1 B_1} \otimes \sigma_{A_2 B_2}) = \mathcal{C}^h_{\bar{A}B} (\rho_{A_1 B_1}) +\mathcal{C}^h_{\bar{A}B} (\sigma_{A_2 B_2}).
\end{equation}
This is (\ref{lemma2-11}).  Similarly we obtain (\ref{lemma2-22}).
The other equalities (\ref{lemma2-33}-\ref{lemma2-66}) follow straightforwardly.
This completes the proof of lemma 2.

\subsection{Proof of proposition 3}

For the CQ states in (\ref{CQdefine}), the optimum POVM on A in the definition of $\mathcal{C}_{\bar{A}\bar{B}}$ is actually the projection onto the basis states $\{\left|i \right\rangle _A \}$ \cite{wu2009correl}.
Therefore, $\mathcal{C}_{\bar{A}\bar{B}}$ is the maximum value of mutual information for the joint probability distribution $\{ p_{ij} \}$ over all possible choices of local POVM $\{\Pi_j^B \}$,
\begin{equation}
\mathcal{C}_{\bar{A}\bar{B}}= \max_{\{\Pi_j^B\}} I(A:B) (p_{ij}) .
\end{equation}
with
\begin{equation}
p_{ij}= p_i tr_B( \Pi_j^B \rho_i^B) .
\end{equation}
On the other hand,
$\mathcal{C}^h_{A\bar{B}} =\max_{\Pi_j^B}\{ S(\rho_A) -\sum_j q_j S(\rho_j^A) \}$, where
$q_j = \sum_i p_i tr(\Pi_j^B \rho_i^B) =\sum_i p_{ij}$ and $\rho_j^A =\sum_i p_{i|j} \left|i \right\rangle_A \left\langle i \right|$ with $p_{i|j}=p_{ij}/q_j$. One easily has $S(\rho_A) = - \sum_i p_i \log_2 p_i \equiv H\{p_i ; i\}$, and $S(\rho_j^A) =  -\sum_i p_{i|j} \log_2 p_{i|j} \equiv H\{ p_{i|j}; i\}$.
Thus,
\begin{eqnarray}
\mathcal{C}^h_{A\bar{B}} &=& \max_{\Pi_j^B}\{ H\{p_i; i\} -\sum_j q_j H\{p_{i|j}; i\} \}  \\
               &=& \max_{\Pi_j^B}I(A:B) (p_{ij})  \\
               &=& \mathcal{C}_{\bar{A}\bar{B}} .
\end{eqnarray}
Therefore, for the CQ states, $\mathcal{C}^h_{A\bar{B}} =\mathcal{C}_{\bar{A}\bar{B}}$, and $\mathcal{Q}^d_{A\bar{B}} =\mathcal{Q}^s_{AB}$.
The optimum POVM that achieves $\mathcal{C}_{\bar{A}\bar{B}}$ actually involves a local projective measurement (by Alice) onto the basis states $\left|i \right\rangle_A$ and a POVM (by Bob) which is optimum to achieve $\mathcal{C}^h_{A\bar{B}}$.
From lemma 2, we know that $\mathcal{C}^h_{A\bar{B}}$ and $\mathcal{Q}^d_{A\bar{B}}$ are additive for the CQ states, therefore, $\mathcal{C}_{\bar{A}\bar{B}}$ and $\mathcal{Q}^s_{AB}$ are also additive for the CQ states. Hence, the last two equalities in proposition 3 is proved.  The proof of proposition 3 is completed.

\subsection{Proof of proposition 4}

The proof is straightforward.
Statement (i) in proposition 4 follows from (\ref{EFCrelationregularized}) and the definition of the coherent information $I^C$.
Statement (ii) follows from statement (i) and (\ref{EDIrelation}).
When the regularized measure of classical correlation $\overline{\mathcal{C}}^h_{A\bar{C}}$ reaches its upper
bound $\mathcal{S}^m_{AC}$, i.e., $\overline{\mathcal{C}}^h_{A\bar{C}} = \mathcal{S}^m_{AC}$, we have $E^C(\rho_{AB}) -E^D(\rho_{AB}) \leq 0$ from (\ref{ent2}). Since $E^C(\rho_{AB}) -E^D(\rho_{AB}) \geq 0$, we immediately have $E^C(\rho_{AB}) =E^D(\rho_{AB})$. Therefore, statement (iii) is proved.  Statement (iv) is obvious as all the arguments in this proof are still true when each subscript AC is replaced with BC.

\subsection{Proof of (\ref{PhiECED}), (\ref{PhiSm}) and (\ref{PhiChregAC})}

For the tripartite state $\left| \Phi \right\rangle_{ABC}$ in (\ref{triplephi}), the marginal density matrices are given as
\begin{eqnarray}
\rho_{AB} &=&  \varrho_{a_2 b_2}  \otimes \sigma_{a_1 a_3 b_1 b_3} \\
\rho_{AC} &=&  \varrho_{a_3 c_2}  \otimes \sigma_{a_1 a_2 c_1 c_3} \\
\varrho_{a_2 b_2} &=& \left| EPR \right\rangle_{a_2 b_2} \left\langle EPR \right| \\
\varrho_{a_3 c_2} &=& \left| EPR \right\rangle_{a_3 c_2} \left\langle EPR \right| \\
\sigma_{a_1 a_3 b_1 b_3} &=& \frac{1}{2} (\left| 00\right\rangle \left\langle 00 \right| +\left| 00\right\rangle \left\langle 11 \right|)_{a_1 b_1}  \otimes \frac{1}{4} I_{a_3 b_3} \\
\sigma_{a_1 a_2 c_1 c_3} &=& \frac{1}{2} (\left| 00\right\rangle \left\langle 00 \right| +\left| 00\right\rangle \left\langle 11 \right|)_{a_1 c_1}  \otimes \frac{1}{4} I_{a_2 c_3} .
\end{eqnarray}

From one copy of $\rho_{AB}$ we can distillate one copy of EPR state via LOCC, and we can create one copy of $\rho_{AB}$ from one copy of EPR state via LOCC as well, therefore, $E^C \leq 1$, $E^D \geq 1$. As entanglement cost is no less than entanglement of distillation, we immediately have (\ref{PhiECED}).
From the expression of $\rho_{AC}$, it is straightforward to get (\ref{PhiSm}).

In order to prove (\ref{PhiChregAC}), we need to calculate $\overline{\mathcal{C}}^h_{A\bar{C}}$. We recall the definition
\begin{eqnarray}
\overline{\mathcal{C}}^h_{A\bar{C}} &=& \lim_{n\rightarrow \infty} \frac{1}{n} \mathcal{C}^h_{A\bar{C}} (\rho_{AC}^{\otimes n}) \\
&=& \lim_{n\rightarrow \infty} \frac{1}{n} \mathcal{C}^h_{A\bar{C}} (\varrho_{a_3 c_2}^{\otimes n}  \otimes \sigma_{a_1 a_2 c_1 c_3}^{\otimes n}) .   \label{appe-11}
\end{eqnarray}
Since $\varrho_{a_3 c_2}^{\otimes n}$ is a pure state, we have $\mathcal{C}^h_{A\bar{C}}( \varrho_{a_3 c_2}^{\otimes n} )= S(\varrho_{a_3}^{\otimes n}) = n$.  It is straightforward to get $\mathcal{S}^m (\sigma_{a_1 a_2 c_1 c_3}) = 1$, this upper bound can be achieved by $\mathcal{C}^h_{A\bar{C}}$ when Camilla projects her systems $c_1c_3$ onto the computational basis, therefore $\mathcal{C}^h_{A\bar{C}}(\sigma_{a_1 a_2 c_1 c_3})=1$. Since $\sigma_{a_1 a_2 c_1 c_3}$ is a separable state, from lemma 2, we have
$\mathcal{C}^h_{A\bar{C}}(\sigma_{a_1 a_2 c_1 c_3}^{\otimes n})=n$, and
\begin{eqnarray}
&& \mathcal{C}^h_{A\bar{C}}(\varrho_{a_3 c_2}^{\otimes n}  \otimes \sigma_{a_1 a_2 c_1 c_3}^{\otimes n}) \nonumber  \\
&=& \mathcal{C}^h_{A\bar{C}}(\varrho_{a_3 c_2}^{\otimes n} ) +\mathcal{C}^h_{A\bar{C}}(\sigma_{a_1 a_2 c_1 c_3}^{\otimes n})  \nonumber  \\
&=& n +n =2n .  \label{appe-22}
\end{eqnarray}
Thus, (\ref{PhiChregAC}) directly follows from (\ref{appe-11}) and (\ref{appe-22}). This completes the proof of (\ref{PhiECED}), (\ref{PhiSm}) and (\ref{PhiChregAC}).


\begin{thebibliography}{0}

\bibitem{Sc}  E. Schr\"odinger, Naturwissenschaften {\bf 23, }807 (1935).

\bibitem{EPR}  A. Einstein, B. Podolsky, and N. Rosen, Phys. Rev. \textbf{47}, 777 (1935).

\bibitem{BBPS96} C.H. Bennett, H.J. Bernstein, S. Popescu, and B. Schumacher, Phys. Rev. A {\bf 53}, 2046 (1996).

\bibitem{Bennett96hash} C.H. Bennett et al, Phys. Rev. A {\bf 54}, 3824 (1996).

\bibitem{VPRK97} V. Vedral, et al, Phys. Rev. Lett. {\bf 78}, 2275 (1997).

\bibitem{CW04} M. Christandl and A. Winter, J. Math. Phys. {\bf 45}, 829 (2004).

\bibitem{OZ01} H. Ollivier and W. H. Zurek, Phys. Rev. Lett. {\bf 88}, 017901 (2001).

\bibitem{Oppenheim2002} J. Oppenheim, M. Horodecki, P. Horodecki, and R. Horodecki, Phys. Rev. Lett. {\bf 89}, 180402 (2002).

\bibitem{Horodecki2003a} M. Horodecki, K. Horodecki, P. Horodecki, R. Horodecki, J. Oppenheim, A. Sen De, and U. Sen, Phys. Rev. Lett. {\bf 90}, 100402 (2003).

\bibitem{Luo2008} S. Luo, Phys. Rev. A {\bf 77}, 022301 (2008).

\bibitem{Piani2008} M. Piani, P. Horodecki, and R. Horodecki, Phys. Rev. Lett. {\bf 100}, 090502
(2008).

\bibitem{wu2009correl} S. Wu, U. V. Poulsen, and K. M{\o}lmer,  Phys. Rev. A 80, 032319 (2009).

\bibitem{Modi2010} K. Modi, T. Paterek, W. Son, V. Vedral, and M. Williamson, Phys. Rev. Lett. {\bf 104}, 080501 (2010).

\bibitem{Dakic2010} B. Dakic, V. Vedral, and C. Brukner, Phys. Rev. Lett. {\bf 105}, 190502 (2010).

\bibitem{LuoFu2010} S. Luo, and S. Fu, Phys. Rev. A. {\bf 82}, 034302 (2010).

\bibitem{AD2010} G. Adesso, and A. Datta, Phys. Rev. Lett. {\bf 105}, 030501 (2010).

\bibitem{GP2010} P. Giorda, and M.G.A. Paris, Phys. Rev. Lett. {\bf 105}, 020503 (2010).

\bibitem{Modi2011} K. Modi, A. Brodutch, H. Cable, T. Paterek, and V. Vedral, arxiv: 1112.6238 [quant-ph].

\bibitem{KL98} E. Knill and R. Laflamme, Phys. Rev. Lett. {\bf 81}, 5672 (1998).

\bibitem{DSC08} A. Datta, A. Shaji, and C.M. Caves,
Phys. Rev. Lett. {\bf 100}, 050502 (2008).

\bibitem{HV2001} L. Henderson, and V. Vedral, J. Phys. A {\bf 34}, 6899 (2001).

\bibitem{LCS11} M. D. Lang, C. M. Caves and A. Shaji, arXiv: 1105.4920v2 [quant-ph].

\bibitem{DHLST04} D.P. DiVincenzo, M. Horodecki, D.W. Leung, J.A. Smolin and B.M. Terhal,
Phys. Rev. Lett. {\bf 92}, 067902 (2004).

\bibitem{wu2002} S. Wu, and J. Anandan, Phys. Lett. A {\bf 297}, 4 (2002).

\bibitem{DW08} I. Devetak and A. Winter, arXiv: 0304196v2 [quant-ph].

\bibitem{HHT01} P.M Hayden, M. Horodecki, and B. M Terhal, J. Phys. A: Math. Gen. {\bf 34} 6891 (2001).

\bibitem{KW04} M. Koashi and A. Winter, Phys. Rev. A {\bf 69}, 022309 (2004).

\bibitem{DW04} I. Devetak and A. Winter, Phys. Rev. Lett. {\bf 93}, 080501 (2004).

\bibitem{COF11} M. F. Cornelio, M. C. de Oliveira and F. F. Fanchini, Phys. Rev. Lett. {\bf 107}, 020502 (2011).

\bibitem{horodecki09rev} R. Horodecki et al., Rev. Mod. Phys. {\bf 81}, 865 (2009).

\end{thebibliography}
\end{document}